\documentclass[doublecol]{epl2} 
\usepackage{mathrsfs}
\usepackage{amsmath,amssymb}
\def \beq {\begin{equation}}
\def \edq {\end{equation}}
\def \bes {\begin{subequations}}
\def \eds {\end{subequations}}
\def \beqn {\begin{equation*}}
\def \edqn {\end{equation*}}
\def \nn  {\nonumber}

\def \sm {\sigma}

\title{Emergence of Majorana modes in cylindrical nanowires}
\shorttitle{Emergence of Majorana modes in cylindrical nanowires}

\author{Jong Soo Lim \inst{1} \and Rosa Lopez \inst{1,2}\and Lloren\c{c} Serra\inst{1,2}}
\shortauthor{Jong Soo Lim \etal}

\institute{                    
  \inst{1} Institut de Fisica Interdisciplin\`aria i de Sistemes Complexos IFISC (CSIC-UIB), E-07122 Palma de Mallorca, Spain\\
  \inst{2} Departament de Fisica, Universitat de les Illes Balears, E-07122 Palma de Mallorca, Spain
}

\abstract{We present calculations of Majorana edge modes in cylindrical nanowires
of a semiconductor material with proximity-induced superconductivity. 
We consider a Rashba field along the 
transverse direction and an applied magnetic field in arbitrary orientation.
Our analysis is based on exact numerical diagonalizations for the finite cylinder and on the complex band structure for the semi-infinite one.  
Orbital effects are responsible for a strong anisotropy of the critical field for which the effective gap vanishes. Robust Majorana modes are induced by the parallel field component and we find regimes
with more than one Majorana mode on the same edge. Experimentally, they would manifest as a specific sequence of zero-bias conductances as a function of magnetic field.
In the finite cylinder, a degradation of the Majorana 
modes due
to interference of the two edges
leads to oscillating non zero energies
for large enough fields.}
\begin{document}
\maketitle

\section{Introduction}
Novel states of matter, such as quantum Hall systems 
\cite{Ando75,Klitzing80,Laughlin81,Yennie87}, obey 
topological rules. The advent of topological materials 
\cite{kane2005,Bernevig2006,koning,Hasan2010} was a major
breakthrough towards the achievement of a topological quantum computer.
In these materials, quasiparticles are non Abelian composite fermions 
\cite{Wilczek82,Moore91,Ady2010}, the potential basic units 
for quantum information processing \cite{Nayak2008}. Such
peculiar excitations are identical to their own antiparticles
and they are termed Majorana fermions (MFs) honouring Ettore Majorana
who predicted their existence in 1937. 

Majorana's original 
aim was to explain the nature of neutrinos 
as real solutions of the Dirac equation for relativistic particles
\cite{Kopnin91,Volovik99,Read00,Kitaev01,Wilczek09}.
Notoriously, there is not yet a conclusive clue on the existence of 
MFs as elementary particles but they have been engineered in a 
laboratory as quasiparticles. 
Majorana 
fermions in topological materials
can be described as half-fermionic states, so that a pair of MFs
forms a widely distributed full-fermionic state. A pair of MFs defines a qubit whose 
information is encoded not in the individual particles but in a non local pair 
of identical and neutral entities. Non locality gives a great advantage 
to such pairs, they are immune against local sources of decoherence. 
Majorana fermions are incredibly elusive 
because of their lack of characteristic features.
They are chargeless, spinless and energyless, 
what makes them difficult to detect. 

\begin{figure}
\centering
\includegraphics[width=0.15\textwidth]{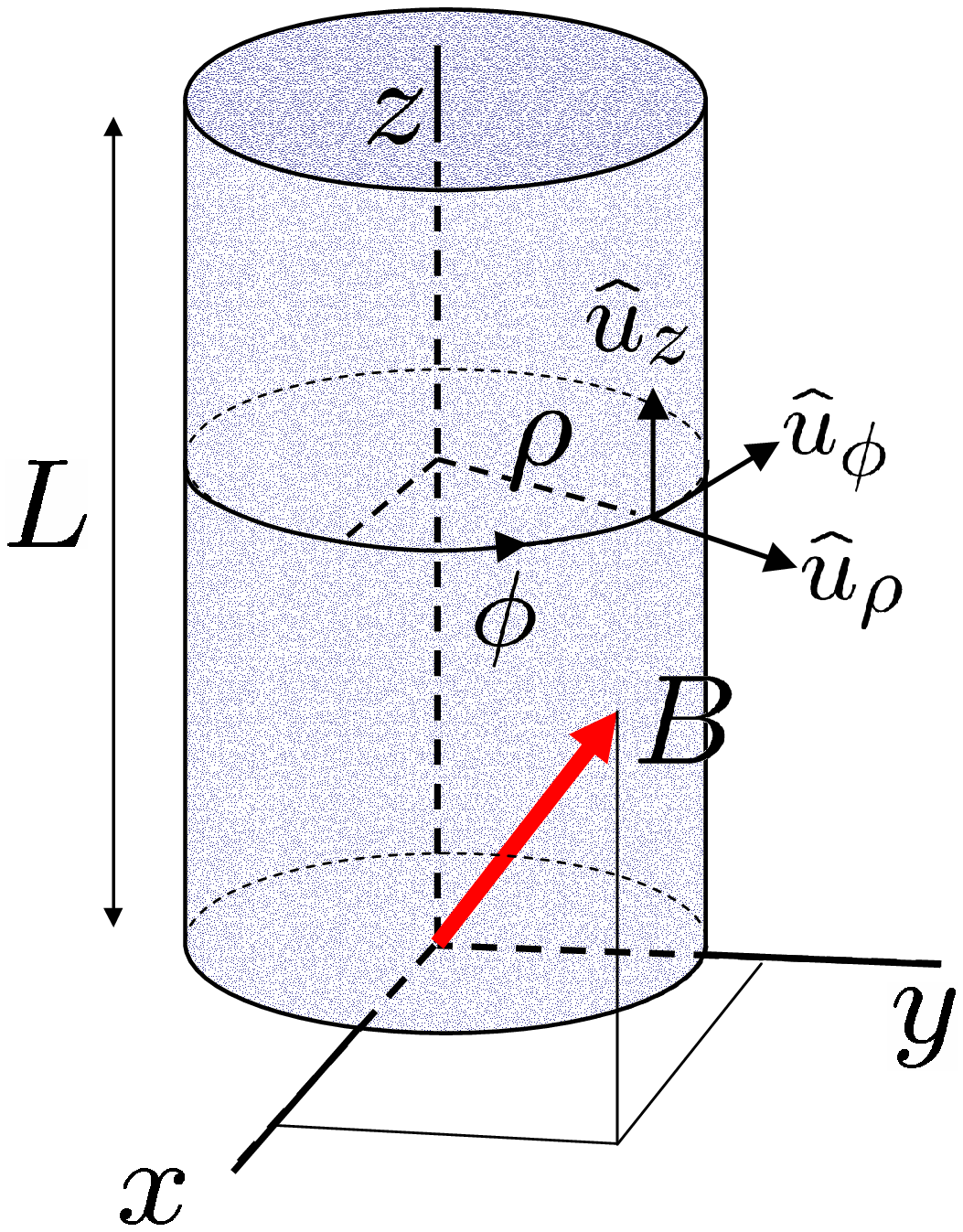}
\caption{
Hollow semiconductor nanowire with cylindrical geometry of length $L$. 
Cylindrical coordinates and unit vectors are indicated.
A magnetic field $\vec{B}$ is applied in an arbitrary direction.}
\label{fig:1}
\end{figure}

The discovery of topological systems  
brought different proposals for the physical 
realization of MFs in solid state devices \cite{Kitaev01,Fu:2008,Beenakker:2012,Alicea:2012}. 
Recently, they have been proposed in semiconductor nanowires 
as a combined effect of
superconductivity, spin-orbit interaction and magnetic field \cite{Lutchyn:2010,Oreg:2010,Alicea:2010a}. 
Mourik and coworkers reported 
in Ref.\ \cite{Mourik:2012} 
signatures of such quasiparticles in long InSb
nanowires ($L\approx 2\mu\rm m$). In the experimental setup an InSb nanowire is connected to a normal-metal gold 
electrode on one side and to a superconducting niobium titanium nitride electrode on the other. 
Since the NbTiN contact is brought in close proximity to the InSb nanowire,
leakage of
Cooper pairs induces superconductivity, at least over a coherence length of a
few hundred nanometers.
InSb is characterized by a large $g$ factor, so that 
the application of an external magnetic field closes the superconductor gap,
shifting some states to zero energy. The spin-orbit interaction  
induces anticrossings 
that separate the zero energy state from the nearest excited states,
preventing the decoherence of the Majorana mode. 
The Majorana-mode evidence is  
a robust zero bias peak
in the tunneling differential
conductance, a zero-bias anomaly (ZBA),
when the nanowire is magnetically driven into the topological phase. 

More experiments have reported evidence of Majorana 
modes in nanowires.
Deng \textit{et} \textit{al}.\ have also observed   
a ZBA in the nonlinear transport measurements with an InSb 
nanowire quantum dot \cite{Deng:2012}. 
Almost simultaneously, Das \textit{et} \textit{al}.\ 
provided similar evidences in shorter hybrid semiconductor/superconductor InAs/Al nanowire 
junctions \cite{Das:2012}. 
Rokhison and coworkers \cite{Rokhison:2012} have presented indirect 
measurements of the fractional a.c.\ Josephson effect in 
hybrid InSb/Nb nanowire 
junctions as a hallmark of the topological matter.  
A more recent experiment by Finck \textit{et} \textit{al}.\ \cite{Finck} 
demonstrates a striking effect: the splitting of the ZBA at high magnetic field
and its oscillatory evolution,
in agreement with the theoretical expectations \cite{rapid,dasSarma}. 

From the theoretical side, most works
dealing with Majorana physics in hybrid semiconductor/superconducting nanowires assume planar geometries. However, all the above mentioned experiments
are done with cylindrical nanowires.  
This motivates us to investigate the theory of Majorana edge modes in 
cylinders.
We quantify the importance of the cylindrical configuration,
rather than a planar one, 
regarding the different
influence of both Rashba field and orbital magnetic effects. 

We find that in cylinders a fixed Rashba direction has to be assumed, 
since a radial one does not
favour the formation of Majoranas.
Orbital effects are inevitably present, however, we demonstrate that they are less dramatic in cylindrical nanowires than in planar ones. In the
latter, a slight deviation of the magnetic field from the nanowire plane 
completely destroys the Majorana state \cite{rapid}.  Remarkably, 
orbital effects drive the cylinder into new topological phases in which 
several Majorana modes can coexist on the same edge. 
We predict the specific sequence of topological phases 
in parallel field for finite and semi-infinite cylinders.
Besides, we show that the critical field
for gap closing strongly depends on the magnetic field orientation due to the 
presence of orbital effects. 

\section{Model}
We study the spectrum of a semiconductor nanowire with cylindrical shape
in a magnetic field. The quasiparticles are assumed to move only on the surface of the cylinder, with a total Hamiltonian 
${\cal H}={\cal H}_{\it kin}+{\cal H}_R+{\cal H}_Z+{\cal H}_{\it pair}$, where the successive 
contributions are kinetic, Rashba spin-orbit, Zeeman, and pairing ones. We include the 
orbital effects of the magnetic field in the kinetic and Rashba Hamiltonians via their dependence 
on quasiparticle momentum. The electron and hole degrees of freedom are represented, as usual,
with Pauli matrices $\tau_{x,y,z}$ acting in the  so-called Nambu space. 

The kinetic energy 
is split into non-magnetic and magnetic terms  
as ${\cal H}_{\it kin}={\cal H}_{\it kin}^{(0)}+{\cal H}_{\it kin}^{(1)}$, where
\begin{eqnarray}
\label{eqkin}
{\cal H}_{\it kin}^{(0)} &=&
\left[\frac{p_{\phi}^2 + p_z^2}{2m^{\ast}} - \mu\right]\tau_z\; ,\\
{\cal H}_{\it kin}^{(1)} &=&
\frac{\hbar^2}{2m^{\ast}} \Biggr[ \frac{\rho}{\ell_z^2}\frac{p_{\phi}}{\hbar}\tau_z + 
\frac{\rho^2}{4\ell_z^4}\tau_z \nn \\ 
&+& 
 2\frac{\rho}{\ell_x^2}\sin\phi\,\frac{p_z}{\hbar}\tau_z
-2\frac{\rho}{\ell_y^2}\cos\phi\, \frac{p_z}{\hbar} 
+\frac{\rho^2}{\ell_x^4}\sin^2{\phi}\, \tau_z \nn\\ 
&+& 
\frac{\rho^2}{\ell_y^4}\cos^2\phi\,\tau_z
-2\frac{\rho^2}{\ell_x^2\ell_y^2} \sin\phi\cos\phi \Biggr]\; .
\end{eqnarray}
We employ the cylindrical coordinate system indicated in Fig.\ \ref{fig:1} and denote 
the chemical potential by $\mu$ and
the magnetic 
length along a general direction $a=x,y,z$ by
$\ell_a=\sqrt{\hbar c/e B_a}$. Notice that, trivially, when
a certain field component 
vanishes the corresponding magnetic length diverges,  giving a vanishing contribution
to ${\cal H}_{\it kin}^{(1)}$. 

The Rashba Hamiltonian originates in an {\em effective} electric field $\vec{\cal E}$ as
${\cal H}_R\propto \vec\sigma\cdot(\vec{p}\times\vec{\cal E}\,)$.
We initially considered a radial field $\vec{\cal E}={\cal E} \hat{u}_\rho$ but, since this
does not lead to the appearance of any Majorana states, we finally assumed a 
uniform direction $\vec{\cal E}={\cal E} \hat{u}_x$.
Similarly to the kinetic term,
we write ${\cal H}_R={\cal H}_R^{(0)}+{\cal H}_R^{(1)}$, with 
\begin{eqnarray}
\label{eqRa}
{\cal H}_R^{(0)} &=& \frac{\alpha}{\hbar} \left[p_z\sm_y\tau_z - \cos\phi\, p_{\phi}\sm_z - \frac{i\hbar}{2\rho}\sin\phi\,\sm_z\right]\; ,
 \\ 
\label{eqRb}
{\cal H}_R^{(1)} &=&
\alpha \left[\frac{\rho}{\ell_x^2}\sin\phi\, \sm_y\tau_z-\frac{\rho}{\ell_y^2}\cos\phi\,\sm_y 
\right.\nn\\
&& \hspace*{2cm}\left. - 
\frac{\rho}{2\ell_z^2}\cos\phi\, \sm_z\right]\,.
\end{eqnarray}

The Zeeman contribution reads
\begin{equation}
{\cal H}_Z=\Delta_B \vec\sigma\cdot\hat{n}\; ,
\end{equation}
where $\hat{n}$ gives the direction of the magnetic field and parameter $\Delta_B$ is related
to the gyromagnetic factor $g$ by $\Delta_B=g \mu_B B /2$. The final Hamiltonian contribution is the 
superconductivity (pairing) one
\begin{equation}
{\cal H}_{\it pair}=\Delta_0 \tau_x\; .
\end{equation}

In summary, the Hamiltonian depends on the field orientation $\hat{n}$, as well as on parameters
$\Delta_0$, $\Delta_B$ and $\alpha$ representing the superconducting, Zeeman and
Rashba coupling strengths, respectively.
In order to investigate realistic setups we assume parameter values 
from Ref.\ \cite{Mourik:2012}. The 
effective mass, Rashba spin-orbit coupling, and radius of InSb nanowire are $m^{\ast} =0.015 m_e$ ($m_e$ is the electron 
bare mass), $\alpha \approx 0.2$  eV{\AA} and $\rho \approx 50$ nm. 
A convenient choice of energy and length units is then
\begin{eqnarray}
E_U &\equiv& \frac{\hbar^2}{m^{\ast}\rho^2} = 2.03\, \rm{meV}\; ,\nn\\
L_U &\equiv& \rho = 50\, {\rm nm}\; .
\end{eqnarray}
We also consider a
superconducting parameter $\Delta_0 \approx 0.25$  meV
and a Zeeman energy that for $g\approx 50$ is related to the field $B$
by $\Delta_{B}= 0.723 B {\rm T}^{-1} E_U$.

\begin{figure}
\centering
\includegraphics[width=0.45\textwidth]{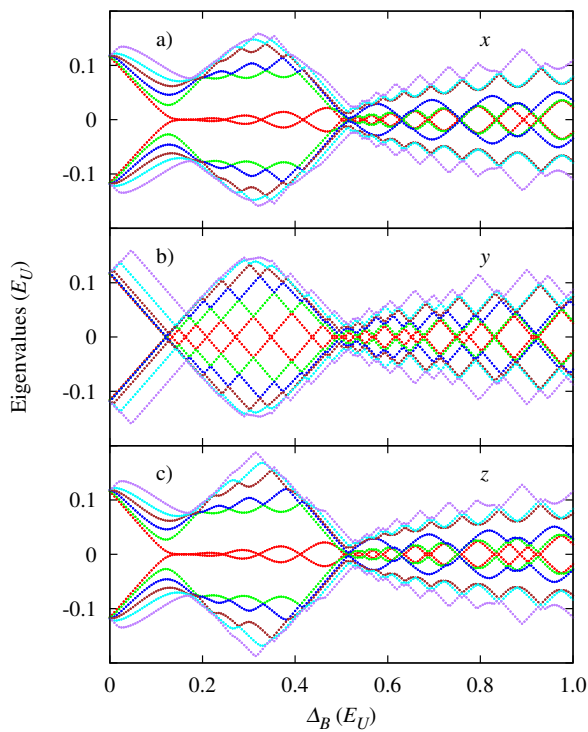}
\caption[Eigenvalues vs $\Delta_B$]{Eigenvalues vs $\Delta_B$ when
 orbital effects are omitted. 
Panels (a), (b) and (c) correspond to applied magnetic fields 
pointing along $x$, $y$ and $z$, respectively. 
$x$ and $z$ directions are almost identical.  
Parameters: $\Delta = 0.12E_U$, $\alpha = 0.2E_UL_U$, $\mu=0E_U$, 
$\rho = L_U$, and $L_z = 30L_U$. Number of basis: $N_n = 51$ and 
$N_m = 31$. Only the twelve eigenvalues closer to zero energy have been 
displayed.}
\label{fig:2}
\end{figure}

\section{Finite cylinders}
We have performed numerical diagonalizations of the Hamiltonian for a cylinder
of length $L$ using as a basis
states $|nms_\sigma s_\tau\rangle$, where $n=1,2,\dots$ represent square well 
eigenstates ($z$ motion), $m=0,\pm 1,\pm2,\dots$ represent $L_z$ angular momentum 
eigenstates ($\phi$ motion) while $s_\sigma=\pm$ and $s_\tau=\pm$ are the usual spin 
and isospin two component eigenstates. 

\begin{figure}
\centering
\includegraphics[width=0.45\textwidth]{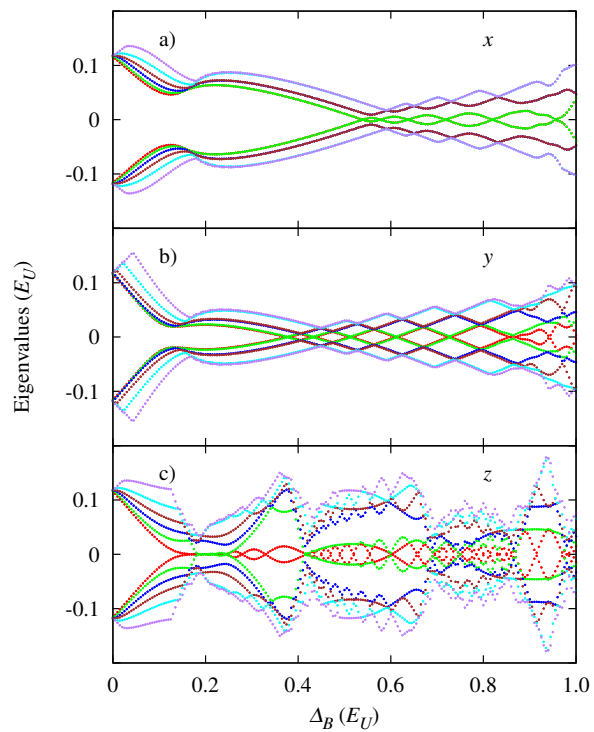}
\caption[Eigenvalues vs $\Delta_B$]{
Same as Fig.\ \ref{fig:2} with orbital 
effects.}
\label{fig:3}
\end{figure}

Figures \ref{fig:2} and \ref{fig:3} show the energy spectrum of a long cylindrical nanowire
($L=30 L_U$)  when the magnetic field is applied along the three Cartesian axis. 
In Fig.\ \ref{fig:2} orbital effects have been neglected ($\ell_{x,y,z}=\infty$)
while in Fig.\ \ref{fig:3} the full Hamiltonian is considered.
For the $\hat{y}$ direction the spectrum in Fig.\ \ref{fig:2}b is dense, with 
accidental crossings at zero energy but no robust Majorana modes.
Field orientations along $\hat{x}$ and $\hat{z}$  (Figs.\ \ref{fig:2}a and \ref{fig:2}c)
are basically equivalent 
and they show the emergence of a zero energy (Majorana) mode beyond a critical field $\Delta^{(c)}_B=0.15 E_U$. Right after the critical field a clear energy gap separates this mode from other excitations. However, as $\Delta_B$ increases an enhanced oscillating behavior manifests the degradation of the Majorana mode.
This is a finite size effect and agrees with our previous study 
of planar geometries \cite{rapid} as well as with Ref.\ \cite{dasSarma}.
Importantly,  recent experimental evidences have confirmed this remarkable 
dependence of the Majorana pair splitting by measurements of the tunneling differential  conductance \cite{Finck}.
When $\Delta_B$ exceeds $0.5E_U$ two more modes emerge in Fig.\ \ref{fig:2}a
and \ref{fig:2}c;  
i.e., a total of three Majorana modes are oscillating 
around zero energy. 
Labeling each topological phase of the cylinder by the number of
different Majorana modes the expected sequence 
when increasing the magnetic field
is then $0-1-3-\dots$. Each $m$-subset contributes 
an independent Majorana after some critical field, while degeneracy of 
$m$ and $-m$ causes modes to emerge in pairs, except for $m=0$. 
We focus next on the changes to this scenario 
induced by orbital effects.

Figure \ref{fig:3} shows the influence of orbital effects on the nanowire energy spectrum. 
Now $\hat{x}$ and $\hat{z}$ orientations are no longer equivalent 
and a large anisotropy in the gap-closing field is obtained.
This critical field is strongly reduced along the sequence $x\to y\to z$.
Along $\hat{x}$ (Fig.\ \ref{fig:3}a)  a state 
of unclear Majorana character 
emerges beyond 
$\Delta_{B}\approx 0.5E_U$,
with an oscillating energy and nearby excited modes. 
For the $\hat{z}$ direction (Fig.\ \ref{fig:3}c) a clear 
Majorana mode emerges at a much 
lower magnetic field, $\Delta_{B}\approx 0.15 E_U$.  
Interestingly, when increasing the field 
along $\hat{z}$
a second Majorana mode seems to be
present for $0.18E_U<\Delta_B<0.27E_U$.
This behavior is in sharp contrast with planar geometries, where Rashba mixing 
precludes the coexistence of more than one Majorana mode.
In cylinders, Rashba mixing is effectively weaker, as its strength 
changes sign with the azimuthal angle [cf.\ Eqs. (\ref{eqRa}) and (\ref{eqRb})].
Notice also that the spin-orbit length, $L_{SO}=\hbar^2/m\alpha\approx 250\; {\rm nm}$,
is $L_{SO}>>\rho$, also indicating a regime of weak spin-orbit 
where multiple Majoranas may coexist \cite{stan:2013}.
Increasing further the magnetic field
in Fig.\ \ref{fig:3}c, oscillations around zero energy become large and
the Majorana character of the states is blurred. 
The sequence of topological phases as the field is increased
suggested in
Fig.\ \ref{fig:3}c is
$0-1-2-1-2-3-2$. 
This is in clear difference with the findings in absence of orbital effects, Fig.\ 
\ref{fig:2}c.

In transport measurements, the presence of multiple Majorana modes affects the 
zero-bias conductance $G_0$. More precisely, $G_0=M{\cal G}$, where $M$ is the 
number of Majorana modes and ${\cal G}$ is a quantization unit
depending on the nature of the contacts (${\cal G}=2e^2/h$ for 
a normal-superconductor junction). 
Although, currently, experiments are still far from this 
limit of high conductances \cite{franz}, the observation of a 
zero bias conductance following the above 
sequence  with magnetic field would be
a clear proof of the cylinder topological phases.

\begin{figure}
\centering
\includegraphics[width=0.40\textwidth]{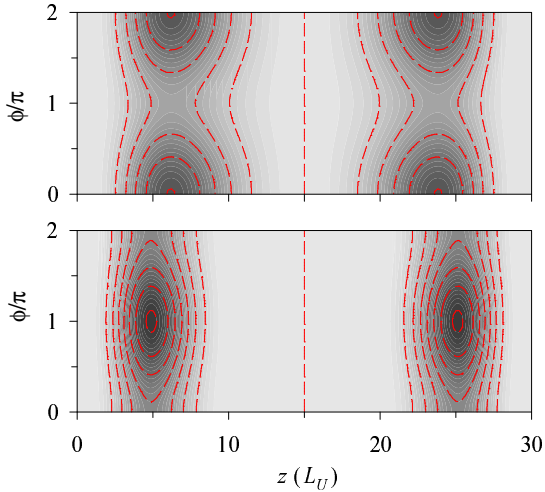}
\caption[figure 5]{
Density distributions corresponding to the two Majorana modes  of
Fig.\ \ref{fig:3}c with $\Delta_B=0.2 E_U$. An arbitrary gray colour scale is used, with dark colour
indicating higher density.}
\label{fig:4n}
\end{figure}

As an illustrative case, Fig.\ \ref{fig:4n} shows the densities of the two Majorana modes
of Fig.\ \ref{fig:3}c when $\Delta_B=0.2 E_U$. As expected, the densities are localized close
to the cylinder edges and decay towards the central part. Remarkably, however, 
the density maxima of 
the two modes take 
complementary azimuthal positions, indicating an effective repulsion between Majorana modes.
We have also obtained that when the mode energies deviate from zero due to the finite size effect, 
for large $\Delta_B$'s, the densities extend more towards the centre, eventually occupying
all of the cylinder length.

\section{Semi-infinite cylinders}
The above results show that finite size effects are relevant for 
an aspect ratio $L/\rho=30$. We have checked that 
for $L/\rho=60$ (results not shown) the finite size oscillations are reduced but 
the behavior is qualitatively the same at high fields. 
Therefore, in order to ascertain the sequence
of topological phases mentioned above, we have directly addressed the semi-infinite 
cylinder using the complex band structure approach of Ref.\ \cite{serra:2013}.  In the 
semi-infinite system 
there is a single edge, the other one lying at infinite distance, and finite-size artifacts are absent by definition.

For a complex wave number $k$ the eigenstate of the 
infinite-cylinder Hamiltonian reads
\begin{equation}
\Psi_k(z,\phi,\eta_\sigma,\eta_\tau)
= e^{ikz} \sum_{s_\sigma s_\tau}{
\psi^{(k)}_{s_\sigma s_\tau}(\phi)\,
\chi_{s_\sigma}(\eta_\sigma)\,
\chi_{s_\tau}(\eta_\tau)
}\; ,
\end{equation} 
where, generically, $\eta_{\sigma,\tau}=\uparrow,\downarrow$ refers to a spin-isospin
discrete variable and $s_{\sigma,\tau}=\pm$ is the corresponding quantum number.
We have determined the allowed wave numbers  $k$
and state amplitudes
$\psi^{(k)}_{s_\sigma s_\tau}$
for a given energy
using the approach of 
Ref.\ \cite{serra:2013}. A Majorana mode is signaled by a zero eigenvalue in matrix
\begin{equation}
{\cal M}_{kk'}=\sum_{s_\sigma s_\tau}{
\int{ d\phi\,
\psi^{(k)*}_{s_\sigma s_\tau}(\phi)\,
\psi^{(k')}_{s_\sigma s_\tau}(\phi)
} }\; .
\end{equation}

In practice the set of complex wave numbers is truncated to a finite size 
using a cut-off in momentum
and the emergence of a zero eigenvalue of matrix ${\cal M}$ is actually seen 
as a steadily decreasing eigenvalue with increasing
value of the cut-off. The approach
of Ref.\ \cite{serra:2013} provides also a method to obtain the critical fields 
corresponding to phase transitions. They are seen as zeros in function ${\cal F}(\Delta_B)$,
the derivative discontinuity at an arbitrary matching  point for $k=0$.

\begin{figure}
\centering
\includegraphics[width=0.45\textwidth]{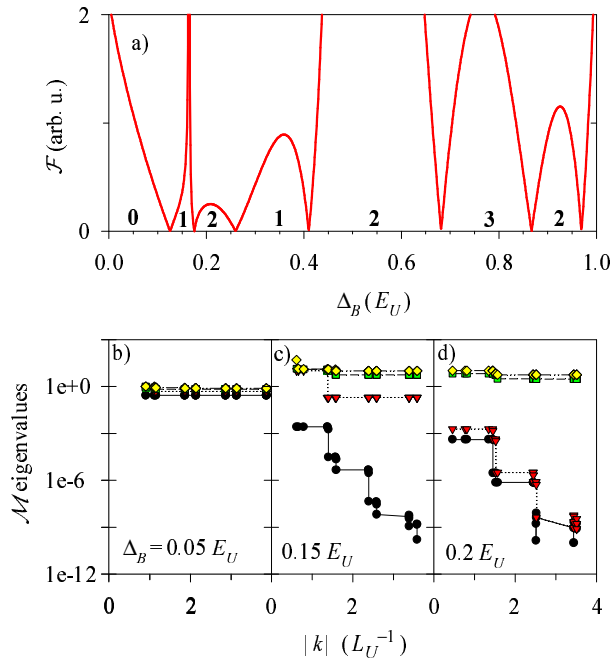}
\caption[figure 4]{
Upper panel: function ${\cal F}$ whose nodes signal the topological transitions.
The number of Majorana edge states are indicated for the low field regions.
Lower panels: Evolution of matrix ${\cal M }$ eigenvalues with the cut-off in
wave numbers (in absolute value). Each eigenvalue tending to zero in the limit
of large $|k|$'s
corresponds to a Majorana mode.
Except for the infinite cylinder length all 
other parameters are the same of Fig.\ \ref{fig:2}c.
}
\label{fig:4}
\end{figure}

Figure \ref{fig:4} confirms the scenario inferred above from the
finite cylinder calculations. Remarkably, the sequence of nodes 
in ${\cal F}$ (Fig.\ \ref{fig:4}a) has a 
clear correspondence with the changes in the finite cylinder spectrum of
Fig.\ \ref{fig:3}c. In the semi-infinite cylinder the critical fields for topological transitions
can 
be determined accurately, while in the finite cylinder transitions are 
blurred, specially at high fields. Lower panels in Fig.\ \ref{fig:4} contain the 
evolution with cut-off in wave numbers of the eigenvalues of matrix
${\cal M}$.  In the first region ($\Delta_B<0.15E_U$) the lowest eigenvalue converges
to $\approx 1$,
indicating that no edge mode is possible. In the second and third regions we see
that one and two eigenvalues, respectively, keep decreasing as the cut-off is increased.
This is a clear evidence that second and third regions contain one
and two Majorana modes, respectively. Similar plots (not shown) 
are obtained for the successive regions, containing the number of Majoranas
indicated in Fig.\ \ref{fig:4}a.
Topological invariance is manifested as independence of the number of Majorana
modes with the precise value of $\Delta_B$, as long as this remains within a given
interval between nodes of Fig.\ \ref{fig:4}a.

\begin{figure}
\centering
\includegraphics[width=0.40\textwidth]{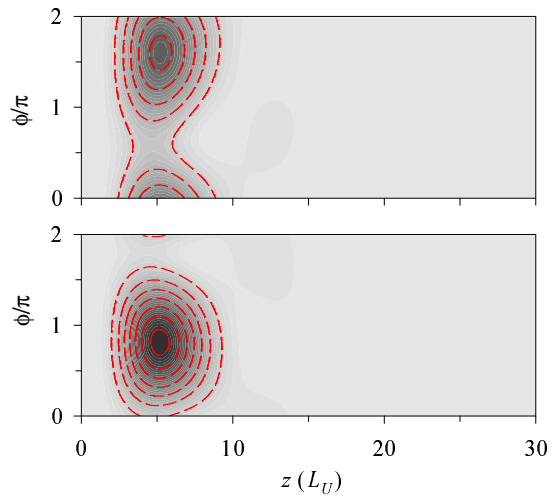}
\caption[figure 5]{
Density distributions corresponding to the two Majorana modes  of
a semi-infinite cylinder with $\Delta_B=0.2 E_U$. 
All other parameters are the same of Fig.\ \ref{fig:2}c.
}
\label{fig:5}
\end{figure}

The spatial densities of the two Majorana modes for
$\Delta_B=0.2E_U$ are shown in Fig.\ \ref{fig:5}.  When compared with the finite cylinder,
densities are in good agreement for cases 
when the  finite-cylinder mode is very close 
to zero energy. If not, large discrepancies  in the bulk of the cylinder
are found. An obvious difference, of course, is that in the finite
cylinder the 
density is repeated on the two edges.
A similar avoided mode overlap is found in both cases, and there is only a
minor angular shift in the relative position of the maxima.
We finally mention that the wave function of Majorana modes
is quasi-real, i.e., its imaginary part almost vanishes.
The wave function is more real the more $k$'s are included in its expansion, and fulfills
$\Psi(z,\phi,\uparrow,\uparrow)=-\Psi(z,\phi,\downarrow,\downarrow)$
and
$\Psi(z,\phi,\uparrow,\downarrow)=\Psi(z,\phi,\downarrow,\uparrow)$. Both things 
are to be expected for proper Majorana solutions.

\section{Conclusions}
In closing, we have investigated the occurrence of 
Majorana
zero-energy
modes at the edges of cylindrical nanowires using 
both finite and semi-infinite models.
We have considered 
a Rashba field along $\hat{x}$ and
a magnetic field in arbitrary direction. Under this situation we  report the strong influence of the orbital effects on
the  appearance of zero energy states. First, 
the critical fields corresponding 
to gap closings
strongly depend on the magnetic field 
orientation. 
Second, in contrast with planar geometries 
we observe phases where
several Majorana modes coexist. Orbital effects define 
the sequence of phases 
with increasing field.
The observation of the sequence $0-1-2-1-2-3\dots$
in the zero bias conductance  
as a function of magnetic field (in units of the conductance quantum)
would be a clear manifestation of the cylinder topological phases.
As in planar geometries, at high fields an oscillating 
behavior signals the degradation of the Majorana 
character due to finite size effects.

\acknowledgments
Work supported by MINECO Grants No.\ FIS2011-23526,
CSD2007-00042 (CPAN),
the Conselleria d'Educaci\'o, Cultura i Universitats
(CAIB) and FEDER.


\begin{thebibliography}{00}

\bibitem{Ando75}
\Name{Ando T., Matsumoto Y. \and Uemura T.}
\REVIEW{J. Phys.\ Soc.\ Jpn.}{39}{1975}{279}.
  
\bibitem{Klitzing80}
\Name{Klitzing K., Dorda D. \and Pepper M.}
\REVIEW{Phys.\ Rev.\ Lett.}{45} (1980){494}.

\bibitem{Laughlin81}
\Name{Laughlin R. B.}
\REVIEW{Phys.\ Rev.\ B}{23}{1981}{5632}.

\bibitem{Yennie87}
\Name{Yennie D. R.}  
\REVIEW{Rev.\ Mod.\ Phys.}{59}{1987}{781}.

\bibitem{kane2005}
\Name{Kane C.L \and Mele E.J}  
\REVIEW{Phys.\ Rev.\ Lett.}{95}{2005}{14680}.

\bibitem{Bernevig2006}
\Name{Bernevig B. A., Taylor L. H. \and  Shou-Cheng Z.}
\REVIEW{Science}{314}{2007}{5806}.

\bibitem{koning} 
\Name{K\"{o}nig J., Wiedmann S.,  Br\"{u}ne C.,  Roth A.,  Buhmann H.,  
Molenkamp L. W., Qi X.-L. \and Zhang S.-C.}
\REVIEW{Science}{318}{2007}{5851}.

\bibitem{Hasan2010}
\Name{Hasan M.  Z. \and Kane L.}
\REVIEW{Rev.\ Mod.\ Phys.} {82}{2010}{3045}.

\bibitem{Wilczek82}
\Name{Wilczek F.} 
\REVIEW{Phys.\ Rev.\ Lett.}{49} {1982}{957}. 

\bibitem{Moore91}
\Name{Moore G. \and Read N.}
\REVIEW{Nuclear Physics B} {360}{1991}{362}.

\bibitem{Ady2010}
\Name{Stern A.}
\REVIEW{Nature} {464}{2010}{187}.

\bibitem{Nayak2008}
\Name{Nayak C., Stern A., Freedman M. \and Das Sarma S.}
\REVIEW{Rev.\  Mod.\  Phys.}{80}{2008}{1083}.

\bibitem{Wilczek09}
\Name{Wilczek F.}
\REVIEW{Nature Phys.} {5}{2009}{614}. 

 \bibitem{Kopnin91}
\Name{Kopnin N. B. \and Salomaa M. M.}
\REVIEW{Phys.\ Rev.\  B}{44}{1991}{9667}. 

 \bibitem{Volovik99}
 \Name{Volovik G.E.}
 \REVIEW{JETP Letters}{70}{1999}{609}. 

\bibitem{Read00}
\Name{Read N. \and Green D.}
\REVIEW{Phys.\  Rev.\  B}{61}{2000}{10267}. 

\bibitem{Kitaev01}
\Name{Kitaev  A. Yu.}
\REVIEW{Physics-Uspekhi}{44}{2001}{131}. 

\bibitem{Beenakker:2012}
\Name{Beenakker C.}
\REVIEW{arXiv:1112.1950}{2012}.

\bibitem{Alicea:2012}
\Name{Alicea J.}
\REVIEW{arXiv:1202.1293}{2012}.

\bibitem{Fu:2008}
\Name{Fu L. \and Kane C. L.}
\REVIEW{Phys.\ Rev.\ Lett.}{100}{2007}{096407}.

\bibitem{Alicea:2010a}
\Name{Alicea J.}
\REVIEW{Phys.\ Rev.\ B}{ 81}{2010}{125318}.

\bibitem{Lutchyn:2010}
\Name{Lutchyn R. M., Sau J. D. \and Das Sarma S.} 
\REVIEW{Phys.\ Rev.\ Lett.} {105}{2010}{077001}.

\bibitem{Oreg:2010}
\Name{Oreg Y., Refael G. \and  Von Oppen F.}
\REVIEW{ Phys.\ Rev.\ Lett.}{105}{2010}{177002}.

\bibitem{Mourik:2012}
\Name{Mourik V., Zuo K., Frolov S. M., Plissard S. R.,  Bakkers E. P. A. M. \and  Kouwenhoven L. P.} 
\REVIEW{Science}{336}{2012}{1003}.

\bibitem{Deng:2012}
\Name{Deng M. T., Yu C. L., Huang G. Y.,  Larsson P., Caroff M. \and  Xu H. Q.} 
\REVIEW{arXiv:1204.4130}{2012}.

\bibitem{Das:2012}
\Name{Das A., Ronen Y., Most Y., Oreg Y., Heiblum M. \and Shtrikman H.}
\REVIEW{arXiv:1205.7073} {2012}.

\bibitem{Rokhison:2012}
\Name{Rokhinson L.P, Liu X. \and Furdyna J. K.}
\REVIEW{Nature Physics} {10}{2012}{1038}.
 
\bibitem{Finck}
\Name{Finck A. D. K.,  Van Harlingen D. J., Mohseni P. K., Jung K. \and Li X.}
\REVIEW{arXiv:1212.1101}{2012}.

\bibitem{rapid} 
\Name{Lim J. S., Serra L.,  L\'opez R. \and Aguado R.}
\REVIEW{Phys.\ Rev.\ B} {86}{2012}{121103}.

\bibitem{dasSarma}
\Name{Das Sarma S., Sau J. D. \and Stanescu T. D.} 
\REVIEW{Phys.\ Rev.\ B}{86}{2012}{220506}.

\bibitem{stan:2013} 
\Name{Stanescu T.D., Lutchyn R.M. \and Das Sarma S.}
\REVIEW{Phys.\ Rev.\ B} {87}{2013}{094518}.

\bibitem{franz}
\Name{Franz M.}
\REVIEW{Nature Nanotech.} {8}{2013}{149}.

\bibitem{serra:2013}
\Name{Serra L.}
\REVIEW{Phys.\ Rev.\ B} {87}{2013}{075440}.

\end{thebibliography}
\end{document}